\input{aipcheck}

\documentclass[
    ,final            % use final for the camera ready runs
  ]
  {aipproc}

\layoutstyle{8x11double}

\usepackage{bm}

\begin{document}

\title{Mode coupling theory for sheared granular liquids}

\classification{<45.70.-n, 61.20.Lc, 64.70.ps, 83.50.Ax, 83.60.Fg>}
\keywords      {<granular matter, liquid theory, mode coupling theory>}

\author{Koshiro Suzuki}{
  address={Canon Inc., 30-2 Shimomaruko 3-chome, Ohta-ku, Tokyo
  146-8501, Japan}
}

\author{Hisao Hayakawa}{
  address={Yukawa Institute for Theoretical Physics, Kyoto University,
  Kitashirakawa Oiwake-cho, Kyoto 606-8502, Japan}
}

\begin{abstract}
Sheared granular liquids are studied by the mode coupling theory.
It is shown that, in contrast to thermostatted systems, current
correlations play an essential role in the dynamics.
The theory predicts that the plateau of the density time-correlator
disappears for most situations, while it appears in the elastic limit.
The result is compatible with molecular dynamics simulations.
\end{abstract}

\maketitle

%%%%%%%%%%%%%%%%%%%%%%%%%%%%%%%%%%%%%%%%%%%%
%% MAINMATTER
%%%%%%%%%%%%%%%%%%%%%%%%%%%%%%%%%%%%%%%%%%%%

\section{Introduction}

\hspace{0.5em}
It is believed in the description of granular flows that the kinetic
theory of Boltzmann-Enskog is applicable up to volume fraction $\varphi
\le 0.5$
%\cite{BP, G2003, JR1985, GD1998, L2005, SH2007}.
\cite{BP, JR1985, GD1998, L2005, SH2007}.
However, any reliable thoery for dense granular flows with $\varphi
> 0.5$ does not exist, where the picture of instantaneous collision
breaks down.
Experiments and simulations suggest that the constitutive equation of
the shear stress exhibits a crossover from the Bagnold law to a
power-law behavior with a non-trivial exponent, and eventually results
in a yield stress for large $\varphi$ \cite{HOS2007, H2008}.
To understand this behavior by a first-principle theory is desired.

On the other hand, Liu and Nagel \cite{LN1998} suggested the existence
of a closed relationship between the jamming transition and the glass
transition.
Since then, to clarify this relation has been one of the hottest
subjects in both granular and glassy physics \cite{IBS2012}.
This situation has motivated us to apply the mode coupling theory (MCT)
\cite{G} successfully used for glassy materials to granular materials
\cite{HO2008, KSZ2010}.
This is in accordance with the establishment of the notion of ``granular
liquids'' 
%
%\cite{K2006, OK2007, ROK2009, K2009, OH2009, OH2009-2}%
\cite{K2006, OK2007, OH2009, OH2009-2}%
, where long-time, long-range correlations are significant, as well as
in molecular liquids.
Hayakawa and Otsuki \cite{HO2008} introduced a MCT for sheared granular
liquids of hard-core inelastic spheres, but realistic granular grains
are soft and the application of the pseudo-Liouvillian \cite{BP} is
unnecessary.
Kranz {\it et al.} \cite{KSZ2010} studied MCT for driven inelastic
spheres with a Gaussian white-noise thermostat, but the correspondence
of their model with the actual vibrating granular systems, which
incorporate non-Gaussian mechanical forces, is quite unclear.

With the above situation in consideration, we attempt to construct a
liquid theory for sheared granular spheres with a soft-core potential,
starting from the Liouville equation %
%
%\cite{BDS1997, DBB2008, BDB2008, GN2000, DB2002, DBL2002, LBD2002,
%BRMMGdS2005, COH2010-3, HCO2010}.
%
\cite{DBB2008, BDB2008, COH2010-3, HCO2010}.
In particular, we apply the MCT to obtain a set of closed equations for
the time-correlators.
Although MCT is still not completely established as a theory of glasses,
its problem is not serious at least for sheared systems, so we adopt the
conventional approach of projection operators and the Mori-type
equations \cite{Mori1965, Z}.
The physical significance of the current correlation is demonstrated by
exhibiting its effect on the slow dynamics.

\section{Formulation}

\hspace{0.5em}
In this section, we present the equation of motions for the
time-correlators (MCT equation).
Detailed derivations for the basic formulation of sheared underdamped
MCT can be found in Ref.~\cite{SH2012}, so we focus on the issues
specific to granular systems.

Our microscopic starting point is the SLLOD equation \cite{EM} for a
system under a uniform and stationary shear with $N$ spherical particles
of mass $m$ and diameter $d$,
\begin{eqnarray}
\dot{\bm{r}}_i(t)
&=&
\frac{\bm{p}_i(t)} {m}
+
\bm{\kappa}\cdot \bm{r}_i(t),
\label{Eq:SLLOD_r}
\\
\dot{\bm{p}}_i(t)
&=&
\bm{F}_i^{(\mathrm{el})}(t)
+
\bm{F}_i^{(\mathrm{dis})}(t)
-
\bm{\kappa}\cdot \bm{p}_i(t),
\label{Eq:SLLOD_p}
\end{eqnarray}
where $\kappa^{\lambda \mu} = \dot{\gamma} \delta^{\lambda x}
\delta^{\mu y}$ is the $\lambda\mu$ component of the shear-rate tensor
($\lambda, \mu = x,y,z$), $\dot{\gamma}$ is the shear rate,
$\bm{F}_i^{(\mathrm{el})}$ is the conservative force with a
soft-core potential, $\bm{F}_i^{(\mathrm{dis})} = \sum_{j\neq i}
\bm{F}_{ij}^{(\mathrm{dis})}$ is the viscous dissipative force, where
$\bm{F}_{ij}^{(\mathrm{dis})}$ is the pairwise contact disspative force,
\begin{eqnarray}
\bm{F}_{ij}^{(\mathrm{dis})}
\hspace{-0.5em}
&=&
\hspace{-0.5em}
- \hat{\bm{r}}_{ij} 
\mathcal{F}(r_{ij})
\left( \bm{v}_{ij} \cdot \hat{\bm{r}}_{ij} \right),
\label{Eq:Fvis}
\\
\mathcal{F}(r_{ij})
\hspace{-0.5em}
&\equiv&
\hspace{-0.5em}
\zeta
\Theta( d - r_{ij}),
\label{Eq:F}
\end{eqnarray}
with $\bm{v}_{ij} \equiv \bm{v}_i - \bm{v}_j$, $\bm{r}_{ij} \equiv
\bm{r}_i - \bm{r}_j$, $r_{ij} \equiv | \bm{r}_{ij}|$,
$\hat{\bm{r}}_{ij} \equiv \bm{r}_{ij} / r_{ij}$, and $\Theta(x)$ is the
Heaviside's step function, where $\Theta(x)=1$ for $x>0$ and $\Theta(x)
= 0$ otherwise.
Here, $\zeta$ is the bare viscous coefficient which is directly related
to the restitution coefficient. 
We consider frictionless granular materials to avoid the complexity to
treat the Coulomb friction and the rotation of grains.
The uniform shear velocity (in the $x$-direction) is $v_{\mathrm{sh}}(y)
= \dot{\gamma} y$, and the distance of the two shear boundaries is
denoted as $L$.
Hence, the velocity at the boundaries may satisfy $v_0^{(\pm)} \equiv
\pm L \dot{\gamma}/2$.
The system is assumed to be in equilibrium before time $t=0$, and
shearing and interparticle dissipation are turned on at $t=0$, after
which a steady-state is asymptotically reached for $t\to\infty$.

It is possible to recast the SLLOD equation, Eqs.~(\ref{Eq:SLLOD_r}) and
(\ref{Eq:SLLOD_p}), into the Liouville equation, and successively by the
application of the projection operator \cite{HCO2010},
%
%\begin{eqnarray}
%\mathcal{P}(t)
%=
%\sum_{\bm{k}} 
%\frac{\left\langle X n_{\bm{k}(t)}^* \right\rangle}{NS_{k(t)}}
%n_{\bm{k}(t)}
%+
%\sum_{\bm{k}} 
%\frac{\left\langle X j_{\bm{k}(t)}^{\lambda *} \right\rangle}{Nv_T^2}
%j_{\bm{k}(t)}^\lambda,
%\end{eqnarray}
%
$\mathcal{P}(t)
=
\sum_{\bm{k}} 
\left\langle X n_{\bm{k}(t)}^* \right\rangle / \left( NS_{k(t)} \right)
n_{\bm{k}(t)}
+
\sum_{\bm{k}} 
\left\langle X j_{\bm{k}(t)}^{\lambda *} \right\rangle / \left( Nv_T^2 \right)
j_{\bm{k}(t)}^\lambda$,
to derive the Mori-type equation for the time-correlators.
Here, $n_{\bm{q}} \equiv \sum_{i=1}^N e^{i\bm{q}\cdot\bm{r}_i} - N
\delta_{\bm{q}, 0}$ and $j_{\bm{q}}^\lambda \equiv \sum_{i=1}^N
p_i^\lambda e^{i\bm{q}\cdot\bm{r}_i} / m$ are the density and
current-density fluctuations, respectively, with $\bm{q}(t) \equiv (q_x,
q_y - \dot{\gamma} t q_x, q_z)$, and $v_T \equiv \sqrt{T/m}$ is the
thermal velocity at the initial equilibrium state.
In granular liquids, it is essential to include the projection onto the
density-current modes $\mathcal{P}_{nj}(t)$, where
\begin{eqnarray}
\mathcal{P}_{nj}(t) X
\equiv
\sum_{\bm{k}>\bm{p}} 
\frac{\left\langle X n_{\bm{k}(t)}^* j_{\bm{p}(t)}^{\lambda *}
      \right\rangle}{N^2 S_{k(t)}v_T^2}
n_{\bm{k}(t)} j_{\bm{p}(t)}^\lambda,
\label{Eq:Pnj}
\end{eqnarray}
to the second projection operator $\mathcal{P}_2(t)$ as \cite{CSOH2012}
\begin{eqnarray}
\mathcal{P}_2(t)
=
\mathcal{P}_{nn}(t)
+
\mathcal{P}_{nj}(t)
\label{Eq:P2}
\end{eqnarray}
to obtain a closure for the Mori-type equation, where
%$\mathcal{P}_{nn}(t)$, where
%%
%\begin{eqnarray}
%\mathcal{P}_{nn}(t) X
%\equiv
%\sum_{\bm{k}>\bm{p}} 
%\frac{\left\langle X n_{\bm{k}(t)}^* n_{\bm{p}(t)}^* \right\rangle}{N^2 S_{k(t)}S_{p(t)}}
%n_{\bm{k}(t)} n_{\bm{p}(t)} ,
%\end{eqnarray}
%
$\mathcal{P}_{nn}(t) X
\equiv
\sum_{\bm{k}>\bm{p}} 
\left\langle X n_{\bm{k}(t)}^* n_{\bm{p}(t)}^* \right\rangle
 / \left( N^2 S_{k(t)}S_{p(t)} \right) 
n_{\bm{k}(t)} n_{\bm{p}(t)}$
is the conventional projection onto the pair-density modes.
This choice of $\mathcal{P}_2(t)$ requires us to consider, in addition
to the conventional time-correlators $\Phi_{\bm{q}}(t) \equiv
\left\langle n_{\bm{q}(t)}(t) n_{\bm{q}}^* \right\rangle / N$ and
$H_{\bm{q}}^\lambda(t) \equiv i \left\langle j_{\bm{q}(t)}^\lambda(t)
n_{\bm{q}}^* \right\rangle / N$, two additional
time-correlators,
\begin{eqnarray}
\bar{H}_{\bm{q}}^\lambda (t)
\hspace{-0.5em}
&\equiv&
\hspace{-0.5em}
\frac{i}{N}
\left\langle n_{\bm{q}(t)}(t) j_{\bm{q}}^{\lambda *} \right\rangle,
\label{Eq:bH}
\\
C_{\bm{q}}^{\lambda \mu}(t)
\hspace{-0.5em}
&\equiv&
\hspace{-0.5em}
\frac{1}{N}
\left\langle j_{\bm{q}(t)}^\lambda(t) j_{\bm{q}}^{\mu *} \right\rangle.
\label{Eq:C}
\end{eqnarray}
These time-correlators turn out to play essential and significant roles
in granular liquids.

\subsection{Isotropic approximation}

It is almost hopeless to deal with four time-correlators with tensor
indices.
Hence, we attempt to reduce the degrees of freedom by applying the
isotropic approximation.
Since dissipation occurs irrespective of the anisotropy of the strain,
application of this approximation is expected not to be fatal.

Based on the isotropic approximation, it is able to reduce the
vector/tensor time-correlators into two scalar time-correlators, $\{
\Phi_q(t), \Psi_q(t)\}$.
In particular, $H_{\bm{q}}^\lambda(t)$ is reduced to
$H_{\bm{q}}^\lambda(t) \simeq q^\lambda(t) \frac{d}{dt} \Phi_{q}(t) /
q(t)^2$ as usual \cite{SH2012}, and $\bar{H}_{\bm{q}}^\lambda(t)$ to
\cite{SH2012-2}
%%
%\begin{eqnarray}
%H_{\bm{q}}^\lambda(t)
%\simeq
%\frac{q^\lambda(t)}{q(t)^2}
%\frac{d}{dt}
%\Phi_{q}(t),
%\end{eqnarray}
%%
%
\begin{eqnarray}
\bar{H}_{\bm{q}}^\lambda(t)
\simeq
- q^\lambda(t) \Psi_{q} (t).
\end{eqnarray}
The correlator $C_{\bm{q}}^{\lambda\mu}(t)$ is reduced to
$C_{\bm{q}}^{\lambda\mu}(t) \simeq C_q(t) \delta^{\lambda\mu}$
\cite{SH2012-2}, where $C_q(t)$ is written in terms of $\Psi_q(t)$ as
\begin{eqnarray}
C_q(t)
=
\frac{d}{dt}
\Psi_q(t)
+
\frac{1}{2}
\frac{\Psi_q(t)}{q(t)^2}
\frac{d}{dt}
q(t)^2.
\end{eqnarray}
The resulting MCT equations in the weak-shear regime read
\begin{eqnarray}
\frac{d^2}{dt^2}
\Phi_q(t)
\hspace{-0.5em}
&=&
\hspace{-0.5em}
- v_T^2 
\frac{q(t)^2}{S_{q(t)}} \Phi_q(t)
-
A_{\bm{q}(t)}
\frac{d}{dt}
\Phi_{q}(t)
\nonumber \\
\hspace{-0.5em}
&&
\hspace{-0.5em}
- \int_0^t ds
\bar{M}_{\bm{q}(s)}(t-s)
\frac{d}{ds}
\Phi_{q}(s),
\\
\frac{d^2}{dt^2}
\Psi_q(t)
\hspace{-0.5em}
&=&
\hspace{-0.5em}
- \frac{v_T^2 }{3}
\frac{q(t)^2}{S_{q(t)}} \Psi_q(t)
-
\frac{1}{3}
A_{\bm{q}(t)}^{\lambda\lambda}
\frac{d}{dt}
\Psi_{q}(t)
\nonumber \\
\hspace{-0.5em}
&&
\hspace{-0.5em}
- \frac{1}{3} \int_0^t ds
\bar{M}_{\bm{q}(s)}^{\lambda\lambda}(t-s)
\frac{d}{ds}
\Psi_{q}(s),
\end{eqnarray}
where
\begin{eqnarray}
A_{\bm{q}} 
=
\frac{4\pi}{3}
\frac{n}{m}
\int_0^\infty dr
r^2 g(r) \mathcal{F}(r)
\left[
1 - j_0(qr) + 2 j_2(qr)
\right],
\label{Eq:A}
\end{eqnarray}
with the spherical Bessel function $j_n(x)$ ($n=0,2$), and
\begin{eqnarray}
A_{\bm{q}}^{\lambda\lambda} 
=
4\pi
\frac{n}{m}
\int_0^\infty dr
r^2 g(r) \mathcal{F}(r)
\left[
1 - j_0(qr)
\right]
\label{Eq:All}
\end{eqnarray}
are the effective viscous coefficients.
Here, $n$ is the average density.
In the weak-shear regime, we keep terms up to linear order in the shear
rate $\dot{\gamma}$ and the bare viscous coefficient $\zeta$, and
neglect quadratic and higher terms.
For instance, $C_q(t)$ is approximated as $C_q(t) \simeq d
\Psi_q(t)/dt$.
\begin{figure}[thb]
\includegraphics[width=8.5cm]{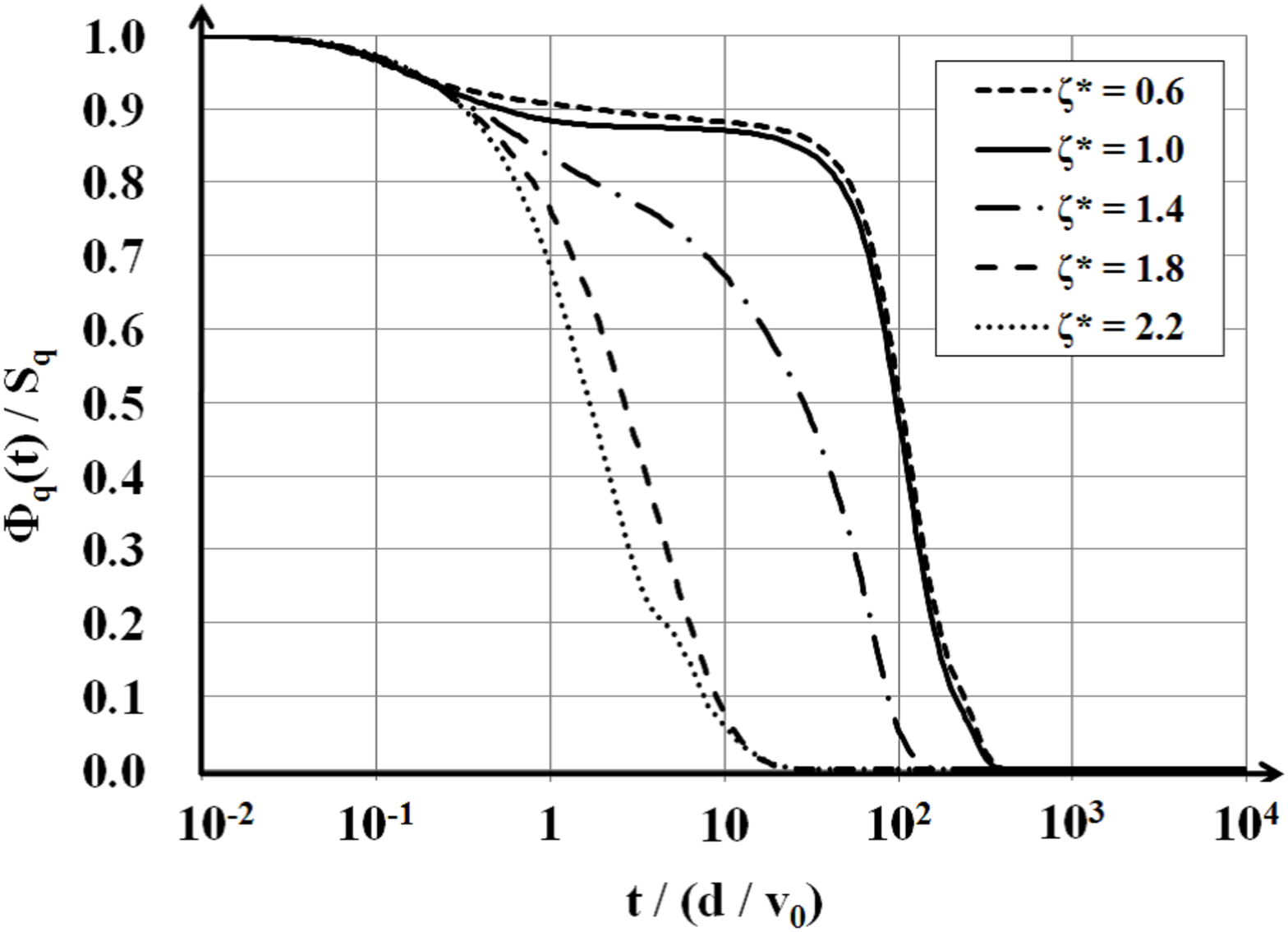} 
\includegraphics[width=8.5cm]{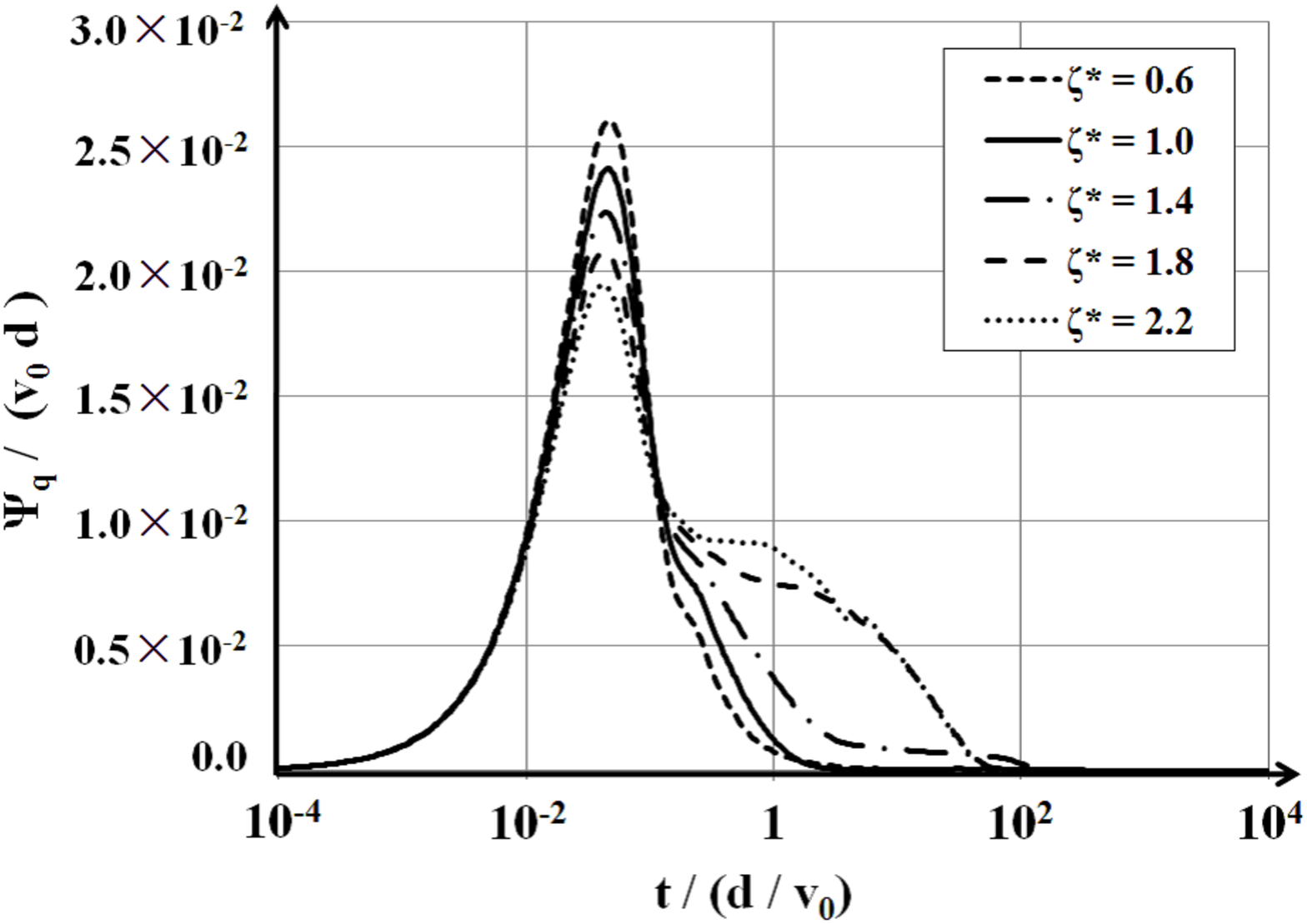} 
\caption{Numerical results for the density time-correlator (left) and
 the density-current time-correlator (right).}
\label{Fig:PhiPsi}
\end{figure}
Note that $\mathcal{F}(r)$ incorporates $\zeta$, cf. Eq.~(\ref{Eq:F}).
According to the choice of $\mathcal{P}_2(t)$ in Eq.~(\ref{Eq:P2}),
there appear four terms in the memory kernels $\bar{M}_{\bm{q}}(\tau)$
and $\bar{M}_{\bm{q}}^{\lambda\lambda}(\tau)$, respectively.
A complete description of these terms is rather lengthy, so we report it
elsewhere \cite{CSOH2012}, and only show the explicit forms of
$\bar{M}_{\bm{q}}(\tau)$ for illustration.
Among the four terms in $\bar{M}_{\bm{q}}(\tau)$,
\begin{eqnarray}
\bar{M}_{\bm{q}}(\tau)
=
\sum_{i=1}^4
\bar{M}_{\bm{q}}^{(i)}(\tau),
\end{eqnarray}
the first term $\bar{M}_{\bm{q}}^{(1)}(\tau)$ is the conventional term
quadratic in $\Phi_q(t)$, 
\begin{eqnarray}
\hspace{-1em}
\bar{M}_{\bm{q}}^{(1)}(\tau)
\hspace{-0.5em}
&=&
\hspace{-0.5em}
\frac{n v_T^2}{2 q^2}
\int
\frac{d^3 \bm{k}}{(2\pi)^3}
V_{\bm{q}(\tau), \bm{k}(\tau), \bm{p}(\tau)}^{(\mathrm{el})}
V_{\bm{q}, \bm{k}, \bm{p}}^{(\mathrm{el})}
\Phi_k(\tau)
\Phi_p(\tau),
\hspace{2em}
\end{eqnarray}
which is responsible for the plateau of the
density time-correlator.
The other three terms originate in dissipation, where the two of them
are given by
\begin{eqnarray}
\bar{M}_{\bm{q}}^{(2)}(\tau) 
\hspace{-0.5em}
&\simeq&
\hspace{-0.5em}
\frac{1}{q^2}
\int 
\frac{d^3 \bm{k}}{(2\pi)^3}
\frac{p^\lambda(\tau)}{p(\tau)^2}
V_{\bm{q}(\tau), \bm{k}(\tau), \bm{p}(\tau)}^{(\mathrm{vis})\lambda}
V_{\bm{q}, \bm{k}, \bm{p}}^{(\mathrm{el})}
\nonumber \\
&&
\times
\Phi_k(\tau)
\frac{d}{d\tau}
\Phi_{p}(\tau),
\label{Eq:M2}
\\
\bar{M}_{\bm{q}}^{(3)}(\tau) 
\hspace{-0.5em}
&\simeq&
\hspace{-0.5em}
- \frac{1}{q^2}
\int 
\frac{d^3 \bm{k}}{(2\pi)^3}
V_{\bm{q}(\tau), \bm{k}(\tau), \bm{p}(\tau)}^{(\mathrm{el})}
p^\lambda
V_{\bm{q}, \bm{k}, \bm{p}}^{(\mathrm{vis})\lambda}
\nonumber \\
&&
\times
\Phi_k(\tau)
\Psi_{p}(\tau)
,
\label{Eq:M3}
\end{eqnarray}
and the remaining $\bar{M}_{\bm{q}}^{(4)}(\tau)$ is neglected since it
is quadratic in $\zeta$.
Here, $V_{\bm{q},\bm{k},\bm{p}}^{(\mathrm{el})} \equiv
(\bm{q}\cdot\bm{k})c_k + (\bm{q}\cdot\bm{p})c_p$, with $\bm{p} \equiv
\bm{q} - \bm{k}$ and the direct correlation function $c_k$, is the
conventional vertex function, and
$V_{\bm{q},\bm{k},\bm{p}}^{(\mathrm{vis})}$ is the dissipative vertex
function, whose explicit form is given by
\begin{eqnarray}
V_{\bm{q}, \bm{k}, \bm{p}}^{(\mathrm{vis})\lambda} 
\equiv
\frac{n}{m}
\int d^3 \bm{r}
g(r) \mathcal{F}(r)
\frac{\bm{q}\cdot\bm{r}}{r^2}
r^\lambda
\left(
e^{i\bm{p}\cdot\bm{r}} - e^{i\bm{k}\cdot\bm{r}} 
\right).
\label{Eq:Vvis}
\end{eqnarray}
Note that Eq.~(\ref{Eq:Vvis}) includes $g(r)\mathcal{F}(r)$ in the
integrand, similarly to Eqs.~(\ref{Eq:A}) and (\ref{Eq:All}).
An important property of the dissipative memory kernels is that
$\bar{M}_{\bm{q}}^{(2)}(\tau)$ and $\bar{M}_{\bm{q}}^{(3)}(\tau)$ are of
opposite sign with respect to $\bar{M}_{\bm{q}}^{(1)}(\tau)$. 
This implies that the interparticle dissipation operates to destroy the
plateau of the density time-correlator in the $\beta$-relaxation regime.
These features are commonly seen in the three terms of
$\bar{M}_{\bm{q}}^{\lambda\lambda}(t)$ as well.

\section{Numerical Calculation}

\hspace{0.5em}
Now we present the result of the numerical calculation.
We first describe the calculational conditions, starting with the
non-dimensionalization scheme.
The units of mass, length, and time are chosen to be $m$, $d$, and
$\tau_0 \equiv d/v_0$, respectively, where $v_0 \equiv v_0^{(+)} -
v_0^{(-)} = L \dot{\gamma}$ is the relative velocity between the shear
boundaries.
In these units, the shear rate $\dot{\gamma} = v_0/L$ is
non-dimensionalized as $\dot{\gamma}^* \equiv \dot{\gamma} \tau_0 =
d/L$, which we require to be small enough, i.e. $d/L \ll 1$
\cite{SH2011}.
In the remainder, non-dimensionalized quantities are denoted with
superscript $^*$.

Next, it is necessary to establish the relation between the bare viscous
coefficient $\zeta$ and the repulsive coefficient $e$ to compare our
result with molecular dynamics simulations.
As explained in Ref.~\cite{OHL2010}, $e$ is related to $\zeta^*$ as $e =
\exp \left[ - \zeta^* \tau_c^* \right]$, where
$\tau_c^* \equiv \pi / \sqrt{2k^* - \zeta^{*2} }$ is the
contact duration time of spheres, if the contact force can be
approximated by the linear spring model.
Here, $k^*$ is the elastic spring coefficient of
$\bm{F}^{(\mathrm{el})}$.
In granular liquids, the steady-state temperature is determined by the
balance of the work by shearing and the interparticle dissipation.
The case of interest might be where the steady-state temperature is of
the same order of the initial equilibrium temperature.
In this case, $\dot{\gamma}^*$ and $\zeta^*$ are not independent, 
and must satisfy a certain scaling relation.
This relation is obtained by requiring $\dot{\gamma}^{*2} \simeq 1 -
e^2$, which leads to
\begin{eqnarray}
\zeta^* 
\simeq
\frac{\dot{\gamma}^{*2}}{2\tau_c^*}. 
\end{eqnarray}
We adopt this scaling relation, with $\tau_c^* = 10^{-4}$, which is
determined by the choice of $k^*$.
%
%In this case, $e=0.99$ and $0.999$ corresponds to $(\dot{\gamma}^*,
%\zeta^*) = (0.142, 14.2)$ and $(4.48\times 10^{-2}, 4.48\times
%10^{-1})$, respectively.

In Fig.~\ref{Fig:PhiPsi}, the results for the density time-correlator
$\Phi_q(t)$ and the density-current time-correlator $\Psi_q(t)$ are
shown.
The shear rate is $\dot{\gamma} \tau_0 = 10^{-2}$, the volume fraction
$\varphi$ is $\epsilon \equiv (\varphi - \varphi_c)/\varphi_c =
+10^{-3}$, where $\varphi_c \simeq 0.516$ is the critical MCT transition
point in equilibrium, and $\zeta$ is varied, which corresponds to
varying $e$.
It can be seen that $\Phi_q(t)$ is almost coincident for $\zeta^* \leq
1.0$, which implies that the effect of dissipation is negligible.
However, for $\zeta^* > 1.0$, which corresponds to the case of $e<0.9999$,
the plateau of $\Phi_q(t)$ is destroyed, and $\Phi_q(t)$ converges to
the Debye-type relaxation curve at $\zeta^* \simeq 2.0$, i.e. $e \simeq
0.9998$.
This drastic demolishing of the plateau is mainly caused by the
dissipative memory kernels $\bar{M}_{\bm{q}}^{(2)}(\tau)$ and
$\bar{M}_{\bm{q}}^{(3)}(\tau)$, Eqs.~(\ref{Eq:M2}) and (\ref{Eq:M3}),
which are of the opposite sign compared to
$\bar{M}_{\bm{q}}^{(1)}(\tau)$, as mentioned previously.
Intuitively, this feature can be regarded as the exhaustion of the
``cage'' due to inelastic collisions, which enables the particles to
escape.
Note that the demolishing of the plateau is accomplished at $t
\sim 10 \tau_0$, which is far below the $\alpha$-relaxation time
$\tau_\alpha \sim 10^2 \tau_0$, where the effect of shear becomes
significant.
This may justify the application of the isotropic approximation.

Furthermore, it is remarkable that, complementarily to $\Phi_q(t)$, a
plateau emerges in the tail of $\Psi_q(t)$ for $\zeta^* > 1.0$.
This implies that the density-current correlation is essential to the
demolishing of the plateau of $\Phi_q(t)$, which supports the role of
the dissipative memory kernels, $\bar{M}_{\bm{q}}^{(2)}(\tau)$ and
$\bar{M}_{\bm{q}}^{(3)}(\tau)$.

The obtained result is compatible with the result from molecular
dynamics, which also indicates that the plateau of the density
time-correlator dissapears for $e<0.99$, while it recovers for the
nearly elastic case \cite{CC2009}.
% 

%\vspace{-1em}
\section{Concluding remarks}

\hspace{0.5em}
It is notable that the projection onto the density-current modes,
Eq.~(\ref{Eq:Pnj}), plays an essential role.
This is in contrast to the sheared thermostatted systems, where
the effect of $\mathcal{P}_{nj}(t)$ has been proved to be negligible
\cite{SH2012-2}.
This difference resides in the nature of the dissipative interaction,
i.e. whether it is single-body or two-body.

According to the integration-through-transient scheme of
Ref.~\cite{FC2002}, it is able to derive a formula for the steady-state
shear stress in terms of time-correlators.
Then, we can obtain the constitutive equation for the shear stress,
which is to be compared to the result of molecular dynamics
\cite{HOS2007, H2008}.
This issue, with an analysis near the jamming transition point, will be
published elsewhere \cite{CSOH2012}.

In conclusion, we have successfully formulated a MCT for sheared
granular liquids.
We have demonstrated that the plateau of the density time-correlator
disappears for $e<0.9999$, as observed in molecular dynamics simulations.
This destruction of the plateau results from the effect of the
density-current correlation.

%%%%%%%%%%%%%%%%%%%%%%%%%%%%%%%%%%%%%%%%%%%%%%%%
%% BACKMATTER
%%%%%%%%%%%%%%%%%%%%%%%%%%%%%%%%%%%%%%%%%%%%%%%%

\begin{theacknowledgments}
\hspace{0.5em}
Numerical calculations in this work were carried out at the computer
facilities at the Yukawa Institute and Canon Inc.  
The authors are grateful to S.-H. Chong and M. Otsuki for stimulating
discussions and careful reading of the manuscript.
\end{theacknowledgments}


\vspace{-1em}
\bibliographystyle{aipproc}   % if natbib is available

\bibliography{Suzuki-Hayakawa}

\providecommand{\noopsort}[1]{}\providecommand{\singleletter}[1]{#1}%
\begin{thebibliography}{30}
\expandafter\ifx\csname natexlab\endcsname\relax\def\natexlab#1{#1}\fi
\providecommand{\enquote}[1]{``#1''}
\expandafter\ifx\csname url\endcsname\relax
  \def\url#1{\texttt{#1}}\fi
\expandafter\ifx\csname urlprefix\endcsname\relax\def\urlprefix{URL }\fi
\providecommand{\eprint}[2][]{\url{#2}}

\bibitem[Brilliantov and Poschel(2010)]{BP}
N.~V. Brilliantov, and T.~Poschel, \emph{Kinetic Theory of Granular Gases},
  Oxford, 2010.

\bibitem[Jenkins and Richman(1985)]{JR1985}
J.~T. Jenkins, and M.~W. Richman, \emph{Phys. Fluids} \textbf{28}, 3485 (1985).

\bibitem[Garzo and Dufty(1998)]{GD1998}
V.~Garzo, and J.~W. Dufty, \emph{Phys. Rev. E} \textbf{59}, 5895 (1998).

\bibitem[Lutsko(2005)]{L2005}
J.~F. Lutsko, \emph{Phys. Rev. E} \textbf{72}, 021306 (2005).

\bibitem[Saitoh and Hayakawa(2007)]{SH2007}
K.~Saitoh, and H.~Hayakawa, \emph{Phys. Rev. E} \textbf{75}, 021302 (2007).

\bibitem[Hatano et~al.(2007)]{HOS2007}
T.~Hatano, M.~Otsuki, and S.~Sasa, \emph{J. Phys. Soc. Jpn.} \textbf{76},
  023001 (2007).

\bibitem[Hatano(2008)]{H2008}
T.~Hatano, \emph{J. Phys. Soc. Jpn.} \textbf{77}, 123002 (2008).

\bibitem[Liu and Nagel(1998)]{LN1998}
A.~J. Liu, and S.~R. Nagel, \emph{Nature} \textbf{396}, 21 (1998).

\bibitem[Ikeda et~al.(2012)]{IBS2012}
A.~Ikeda, L.~Berthier, and P.~Sollich, \emph{Phys. Rev. Lett.} \textbf{109},
  018301 (2012).

\bibitem[G{\"{o}}tze(2009)]{G}
W.~G{\"{o}}tze, \emph{Complex Dynamics of Glass-Forming Liquids. A Mode
  Coupling Theory}, Oxford, 2009.

\bibitem[Hayakawa and Otsuki(2008)]{HO2008}
H.~Hayakawa, and M.~Otsuki, \emph{Prog.\ Theor. Phys.} \textbf{119}, 381
  (2008).

\bibitem[Kranz et~al.(2010)]{KSZ2010}
W.~T. Kranz, M.~Sperl, and A.~Zippelius, \emph{Phys. Rev. Lett.} \textbf{104},
  225701 (2010).

\bibitem[Kumaran(2006)]{K2006}
V.~Kumaran, \emph{Phys. Rev. Lett.} \textbf{96}, 258002 (2006).

\bibitem[Orpe and Kudrolli(2007)]{OK2007}
A.~V. Orpe, and A.~Kudrolli, \emph{Phys. Rev. Lett.} \textbf{98}, 238001
  (2007).

\bibitem[Otsuki and Hayakawa(2009{\natexlab{a}})]{OH2009}
M.~Otsuki, and H.~Hayakawa, \emph{Phys. Rev. E} \textbf{79}, 021502
  (2009{\natexlab{a}}).

\bibitem[Otsuki and Hayakawa(2009{\natexlab{b}})]{OH2009-2}
M.~Otsuki, and H.~Hayakawa, \emph{Eur. Phys. Special Topics} \textbf{179}, 179
  (2009{\natexlab{b}}).

\bibitem[Dufty et~al.(2008)]{DBB2008}
J.~W. Dufty, A.~Baskaran, and J.~J. Brey, \emph{Phys. Rev. E} \textbf{77},
  031310 (2008).

\bibitem[Baskaran et~al.(2008)]{BDB2008}
A.~Baskaran, J.~W. Dufty, and J.~J. Brey, \emph{Phys. Rev. E} \textbf{77},
  031311 (2008).

\bibitem[Chong et~al.(2010)]{COH2010-3}
S.-H. Chong, M.~Otsuki, and H.~Hayakawa, \emph{Phys. Rev. E} \textbf{81},
  041130 (2010).

\bibitem[Hayakawa et~al.(2010)]{HCO2010}
H.~Hayakawa, S.-H. Chong, and M.~Otsuki, \emph{{IUTAM-ISIMM Symposium on
  Mathematical Modeling and Physical Instance of Granular Flows, AIP}}
  \textbf{1227}, 19 (2010).

\bibitem[Mori(1965)]{Mori1965}
H.~Mori, \emph{Prog. Theor. Phys.} \textbf{33}, 423 (1965).

\bibitem[Zwanzig(2001)]{Z}
R.~Zwanzig, \emph{Non-equilibrium statistical mechanics}, Oxford, 2001.

\bibitem[Suzuki and Hayakawa(2013)]{SH2012}
K.~Suzuki, and H.~Hayakawa, \emph{Phys. Rev. E} \textbf{87}, 012304 (2013).

\bibitem[Evans and Morriss(2008)]{EM}
D.~J. Evans, and G.~P. Morriss, \emph{Statistical Mechanics of Nonequilibrium
  Liquids, 2nd ed.}, Cambridge, 2008.

\bibitem[Chong et~al.(????)]{CSOH2012}
S.-H. Chong, K.~Suzuki, M.~Otsuki, and H.~Hayakawa, \ in preparation.

\bibitem[Suzuki and Hayakawa(????)]{SH2012-2}
K.~Suzuki, and H.~Hayakawa, in {\it Slow Dynamics in Complex Systems},
  edited by M. Tokuyama and I. Oppenheim (AIP, New York, 2013).

\bibitem[Saitoh and Hayakawa(2011)]{SH2011}
K.~Saitoh, and H.~Hayakawa, \emph{Gran. Matt.} \textbf{13}, 697 (2011).

\bibitem[Otsuki et~al.(2010)]{OHL2010}
M.~Otsuki, H.~Hayakawa, and S.~Luding, \emph{Prog. Theor. Phys. Suppl.}
  \textbf{184}, 110 (2010).

\bibitem[Ciamarra and Coniglio(2009)]{CC2009}
M.~P. Ciamarra, and A.~Coniglio, \emph{Phys. Rev. Lett.} \textbf{103}, 235701
  (2009).

\bibitem[Fuchs and Cates(2002)]{FC2002}
M.~Fuchs, and M.~E. Cates, \emph{Phys. Rev. Lett.} \textbf{89}, 248304 (2002).

\end{thebibliography}

\IfFileExists{\jobname.bbl}{}
 {\typeout{}
  \typeout{******************************************}
  \typeout{** Please run "bibtex \jobname" to optain}
  \typeout{** the bibliography and then re-run LaTeX}
  \typeout{** twice to fix the references!}
  \typeout{******************************************}
  \typeout{}
 }

\end{document}